\documentclass[aps,pra,reprint,superscriptaddress]{revtex4-2}

\usepackage{amsmath}
\usepackage{amssymb}
\usepackage{graphicx}
\usepackage{hyperref}
\usepackage{blindtext}

\begin{abstract}
	Introducing a spatial chirp into a pulse with a longitudinal vortex, such as a standard pulsed Laguerrre-Gauss beam, results in a vortex pulse with an arbitrary orientation of the line phase singularity between longitudinal and transverse, depending on the amount of chirp. Analytical expressions are given for such pulses with arbitrary topological charge valid at any propagation distance. 
\end{abstract}
\begin{document}

\title{Control of vortex orientation of  ultrashort optical pulses using spatial chirp}
\author{Miguel A. Porras}
\email{miguelangel.porras@upm.es}
\affiliation{Grupo de Sistemas Complejos, ETSIME, Universidad Politécnica de Madrid, Rios Rosas 21, 28003 Madrid, Spain}
\author{Spencer W. Jolly}
\email{spencer.jolly@ulb.be}
\affiliation{Service OPERA-Photonique, Université libre de Bruxelles (ULB), Brussels, Belgium}
\date{\today}
\maketitle

The structuring of ultrashort laser pulse-beams in space or time has long allowed for control of propagation properties, both in linear and nonlinear media, as well as the interaction of the shaped pulses with materials or other physical systems. One of the most well-known examples of shaped or structured light is the optical vortex, whose prototype is the Laguerre-Gaussian (LG) beam solution to the paraxial wave equation, carrying orbital angular momentum (OAM) proportional to the topological charge (or helicity) of the vortex.

Spatiotemporal couplings refer to electric fields that cannot be described as a separable product of complex spatial and temporal functions, or equivalently, spatial and temporal frequency functions~\cite{akturk10} which are increasingly being used to produce new types of laser pulses with interesting properties~\cite{shenY23}. Spatiotemporal optical vortices (STOVs) are a class of spatiotemporal coupled fields featuring a phase line singularity oriented along an axis transverse to the propagation direction~\cite{jhajj16,hancock19,chong20}, e.g., the $y$-direction. The realization of more general STOVs, the phase line singularity of which is oriented at an arbitrary angle between the $y$ and $z$ direction, has been proposed to be possible using photonic crystal structures~\cite{wangH21}. Subsequently, they have been realized experimentally using astigmatic mode converters~\cite{zangY22}. 

In this Letter we provide closed-form, analytical expressions of STOVs with arbitrarily oriented phase line singularities that are solutions of the paraxial wave equation under quasi-monochromatic conditions. These expressions are valid at any propagation distance and for any STOV topological charge. In addition, these expressions represent standard, longitudinal pulsed vortices to which a spatial chirp has been introduced, which provides an alternate and simpler method to realize  experimentally STOVs with arbitrary orientations.  The change of the orientation using spatiotemporal couplings has  been shown  in Ref.~\cite{talposi22}. The analysis of our solution demonstrates that the orientation of the singularity can be finely tuned with the amount of spatial chirp and the pulse and beam parameters. 

Let us start with a LG pulse in frequency domain, of zero radial order, propagating along the $z$ direction, conveniently written as
\begin{equation}\label{LG1}
\hat E_\omega =\hat a_\omega e^{-i(l+1)\psi}\frac{s_0}{s}\left[\frac{\sqrt{2}(x\pm i y)}{s}\right]^{l}e^{\frac{i\omega (x^2+y^2)}{2cq}} e^{i\frac{\omega}{c}z},
\end{equation}
where $l$ is the absolute value of the topological charge, the $\pm$ sign stands for positive and negative topological charge, $\psi=\tan^{-1}(z/z_R)$ is Gouy's phase, $q=z-iz_R$ is the complex beam parameter, $s_0=\sqrt{2z_R c/\omega}$ is the waist Gaussian width, $s=s_0\sqrt{1+(z/z_R)^2}$, and $z_R$ is the Rayleigh distance. For a narrowband spectrum $\hat a_\omega$ about a carrier frequency $\omega_0$, the time-domain field is conveniently written as
\begin{equation}\label{PLG}
E= \frac{1}{2\pi} \int_0^\infty  \hat E_\omega e^{-i\omega t} d\omega \simeq\frac{1}{2\pi}\int_{-\infty}^\infty d\Omega \hat E_\Omega e^{-i\Omega t'}d\Omega \, e^{-i\omega_0 t'},
\end{equation}
where $\Omega=\omega -\omega_0$, $t'=t-z/c$ is the local time,
\begin{equation}\label{LG2}
  \hat E_\Omega =\hat a_\Omega e^{-i(l+1)\psi}\frac{s_0}{s}\left[\frac{\sqrt{2}(x\pm i y)}{s}\right]^{l}e^{\frac{i\omega_0 (x^2+y^2)}{2cq}}\,,
\end{equation}
and $\hat a_\Omega=\hat a_{\omega_0+\Omega}$. We also have used that, for a pulsed vortex with many oscillations, or sufficiently narrow $\hat a_\omega$, the dependence of the beam parameters on frequency can be ignored, thus taking those at the carrier frequency and hence replacing $\omega$ with $\omega_0$ in the last exponential factor in (\ref{LG2}). This approximation is valid in the quasi-monochromatic regime of propagation which yields increasingly accurate results as the number of optical cycles increases well-above the single-cycle pulses. As is well-known, (\ref{LG2}) is a solution of the paraxial wave equation $\partial \hat E_\Omega/\partial z = i(c/2\omega_0)\Delta_\perp \hat E_\Omega $, or $\partial E/\partial z = i(c/2\omega_0)\Delta_\perp E$ in time domain for quasi-monochromatic pulses in non-dispersive media.

\begin{figure}[htb]
	\centering
	\includegraphics[width=86mm]{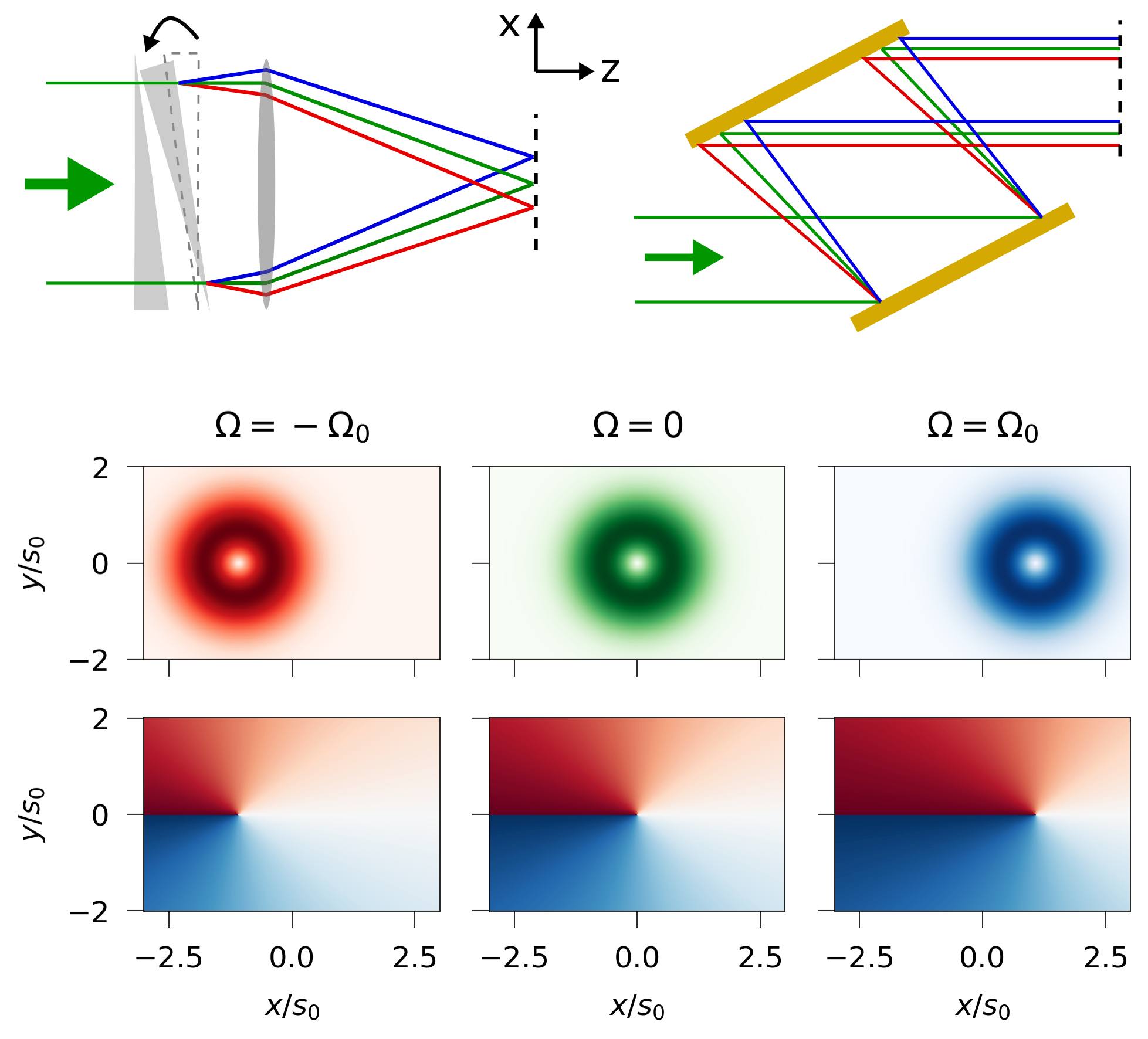}
	\caption{Sketches of how spatial chirp can be done experimentally (top) with either prisms or gratings. The concept of spatial chirp and vortex beams (bottom) with $b=s_0/\Omega_0$, whereby the frequencies at $\pm\Omega_0$ are displaced by $\pm{s_0}$. This affects both the amplitude and the phase, shown here at $z=0$.}
	\label{fig:concept}
\end{figure}

Spatial chirp is a particular spatiotemporal coupling where the different temporal frequencies are separated along one transverse coordinate~\cite{gu04}. Practically, this can be achieved by using dispersive optics such as prisms or gratings in favorable orientations, shown in the top panel of Fig.~\ref{fig:concept}. Here weintroduce the spatial chirp to the separable vortex pulse in (\ref{LG2}), which adds to it important amplitude and phase structure, as in the bottom panel of Fig.~\ref{fig:concept}. Addition of the spatial chirp via transforming $x\rightarrow x-b\Omega$, where $b$ is a constant, leads to
\begin{equation}\label{LG3}
  \hat E_\Omega =\hat a_\Omega e^{-i(l+1)\psi}\frac{s_0}{s}\left[\frac{\sqrt{2}((x-b\Omega)\pm i y)}{s}\right]^{l}e^{\frac{i\omega_0 [(x-b\Omega)^2+y^2]}{2cq}},
\end{equation}
which is non-separable in temporal frequency and space and still satisfies the paraxial wave equation. We have implicitly ignored any other space-time couplings such as angular dispersion to focus on pure spatial chirp. To obtain the non-separable field in space and time field we introduce (\ref{LG3}) into (\ref{PLG}), which readily leads to
\begin{align}\label{PLG2}
\begin{split}
&E= e^{-i(l+1)\psi}\frac{s_0}{s} e^{\frac{i\omega_0 (x^2+y^2)}{2cq}} \left(\frac{\sqrt{2}}{s}\right)^l \\
&\times \frac{1}{2\pi}\int_{-\infty}^\infty \hat a_\Omega (x-b\Omega \pm iy)^l e^{\frac{i\omega_0b^2\Omega^2}{2cq}} e^{-i\Omega t^{\prime\prime}} d\Omega \, e^{-i\omega_0 t'},
\end{split}
\end{align}
where $t^{\prime\prime} = t' +\omega_0bx/cq$. Taking the Gaussian spectrum $\hat a_\Omega  = E_0\sqrt{\pi} (2/\Omega_0)e^{-(\Omega/\Omega_0)^2}$ of Gaussian width $\Omega_0$ (inverse Fourier transform $a(t')= E_0 e^{-t^{\prime 2}/\tau_0^2}$, $\tau_0=2/\Omega_0$), using integral 3.462.4,
\begin{equation}
\int_{-\infty}^\infty \xi^l e^{-(\xi-\beta)^2}d\xi = \int_{-\infty}^\infty (\eta+\beta)^l e^{-\eta^2}d\eta  = \frac{\sqrt{\pi}}{(2i)^l} H_l(i\beta)
\end{equation}
in Ref.~\cite{gradshteyn07} [$H_l(\cdot)$ is the Hermite polynomial of order $l$] after completing the square in the exponential in (\ref{PLG2}) with some changes of variables, we obtain the result
\begin{align}\label{main}
\begin{split}
&E=\frac{E_0}{\alpha\Omega_0}e^{-i(l+1)\psi}\frac{s_0}{s} e^{\frac{i\omega_0 (x^2+y^2)}{2cq}}e^{-\frac{(t'+\omega_0bx/cq)^2}{4\alpha^2}} \left(\frac{\sqrt{2}}{s}\frac{ib}{\alpha}\right)^l\frac{1}{2^l} \\ 
&\times H_l\left\{\left(\frac{\alpha}{ib}\right)  \left[x\left(1+\frac{i\omega_0b^2}{\alpha^2 2cq}\right)\pm iy+i\frac{b}{2\alpha^2}t'\right] \right\} e^{-i\omega_0 t'},
\end{split}
\end{align}
where 
\begin{equation}\label{eq:alpha}
\alpha=\sqrt{\frac{1}{\Omega_0^2}-\frac{i\omega_0 b^2}{2cq}}.
\end{equation}
Equation (\ref{main}) is the main result of this Letter. It represents an optical field with a phase line singularity (for $l=1$) or $l$ phase line singularities oriented at an arbitrary angle in the $z$-$y$ plane, as detailed below. Although the dependence on $z$ in all parameters is omitted for conciseness, (\ref{main}) is valid at any propagation distance.

\begin{figure}[htb]
	\centering
	\includegraphics[width=86mm]{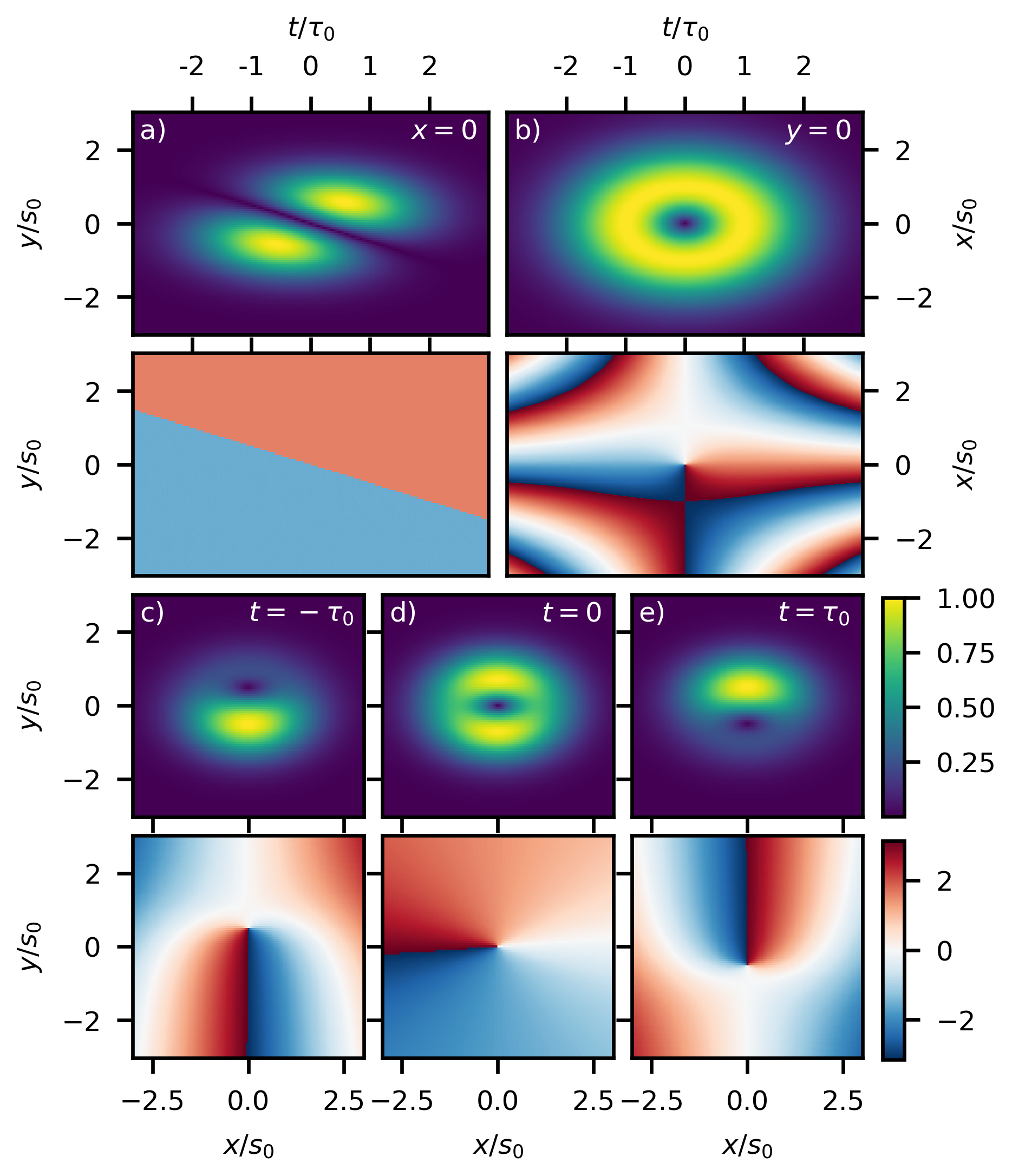}
	\caption{Amplitude and phase of STOV at $z=0$ with $l=1$ and the spatial chirp factor $b=s_0/\Omega_0=s_0\tau_0/2$. Slices are shown with $x=0$ (a), $y=0$ (b), and at $t=\{-\tau_0,0,\tau_0\}$ (c--e), with the amplitude on top and phase on bottom. Each amplitude plot is normalized to its maximum, and the phase is always shown in the range $[-\pi,\pi]$.}
	\label{fig:zeq0}
 %is it amplitude of amplitude what is plotted?
\end{figure}

We first note that $\alpha$ is never zero so that (\ref{main}) does not present singularities. We also note that the factor $1/2^l$ cancels the pre-factor $2^l$ in the highest-power term of the Hermite polynomial. Further, in the limit $b\rightarrow 0$, the factor $(ib/\alpha)^l$ cancels out all terms of the Hermite polynomial except the highest-power term. In addition, $(ib/\alpha)^l$ multiplied by the opposite factor $(\alpha/ib)^l$ in the highest-power term yields unity, and (\ref{main}) reduces to the pulsed Laguerre-Gauss beam
\begin{equation}\label{PLG3}
E =E_0 e^{-\frac{t^{\prime 2}}{\tau_0^2}} e^{-i(l+1)\psi}\frac{s_0}{s}\left[\frac{\sqrt{2}(x\pm i y)}{s}\right]^{l}e^{\frac{i\omega (x^2+y^2)}{2cq}} e^{-i\omega_0t'}.
\end{equation}

With spatial chirp ($b\neq 0$), we first examine the field at $z=0$, where (\ref{main}) simplifies to 
\begin{align}
\begin{split}
&E(z=0)=\frac{E_0}{\alpha\Omega_0}e^{-\frac{(x^2+y^2)}{s_0^2}}e^{-\frac{(t'+2ibx/s_0^2)^2}
{4\alpha^2}}\left(\frac{\sqrt{2}}{s_0}\frac{ib}{\alpha}\right)^{l}\frac{1}{2^l}
\\
&\times H_{l}\left\{\left(\frac{\alpha}{ib}\right)\left[x\left(1-\frac{b^2}{\alpha^2s_0^2}\right)\pm iy+i\frac{b}{2\alpha^2}t'\right]\right\}
e^{-i\omega_0 t'},
\end{split}\label{eq:SC_time_zeq0}
\end{align}
where $\alpha=[(1/\Omega_0)^2+(b/s_0)^2]^{1/2}$ is real. The factor $e^{-(t'+2ibx/s_0^2)^2/4\alpha^2}$ contains the spatial chirp $e^{-ibxt'/s_0^2\alpha^2}$, the Gaussian temporal envelope of enlarged duration $\tau_{0,\rm eff}=\tau_0[1+b^2\Omega_0^2/s_0^2]^{1/2}$, and an anti-Gaussian factor that widens the Gaussian width to the effective width $s_{0,\rm eff}=s_0[1+b^2\Omega_0^2/s_0^2]^{1/2}$. Various slices of the amplitude and phase in the simplest case of $l=1$ and with the relevant value $b=s_0/\Omega_0$ of spatial chirp (see below) are shown in Fig.~\ref{fig:zeq0}, where spatial and spatiotemporal phase singularities can be appreciated. Interestingly, the singularity in the $x=0$ section, observed as a $\pi$-step line $y=\mp (b/2\alpha^2) t'$ in the phase, is seen to move along the $y$ direction with time when the spatial chirp is along $x$. On the contrary, in the $y=0$ section, the amplitude forms a donut with the spatiotemporal phase singularity at the center ---a feature in common with purely transversal STOVs. Transversal slices at different times in the bottom of Fig.~\ref{fig:zeq0} offer an alternate view of the same structure whereby the spatial phase singularity shifts along the $y$ direction, distorting the amplitude pattern over time. At $t'=0$ the amplitude and phase resemble a standard optical vortex slightly elongated along $x$ since the width $s_{0,\rm eff}$ along $x$ is larger than the width $s_0$ along $y$.

\begin{figure}[htb]
	\centering
	\includegraphics[width=86mm]{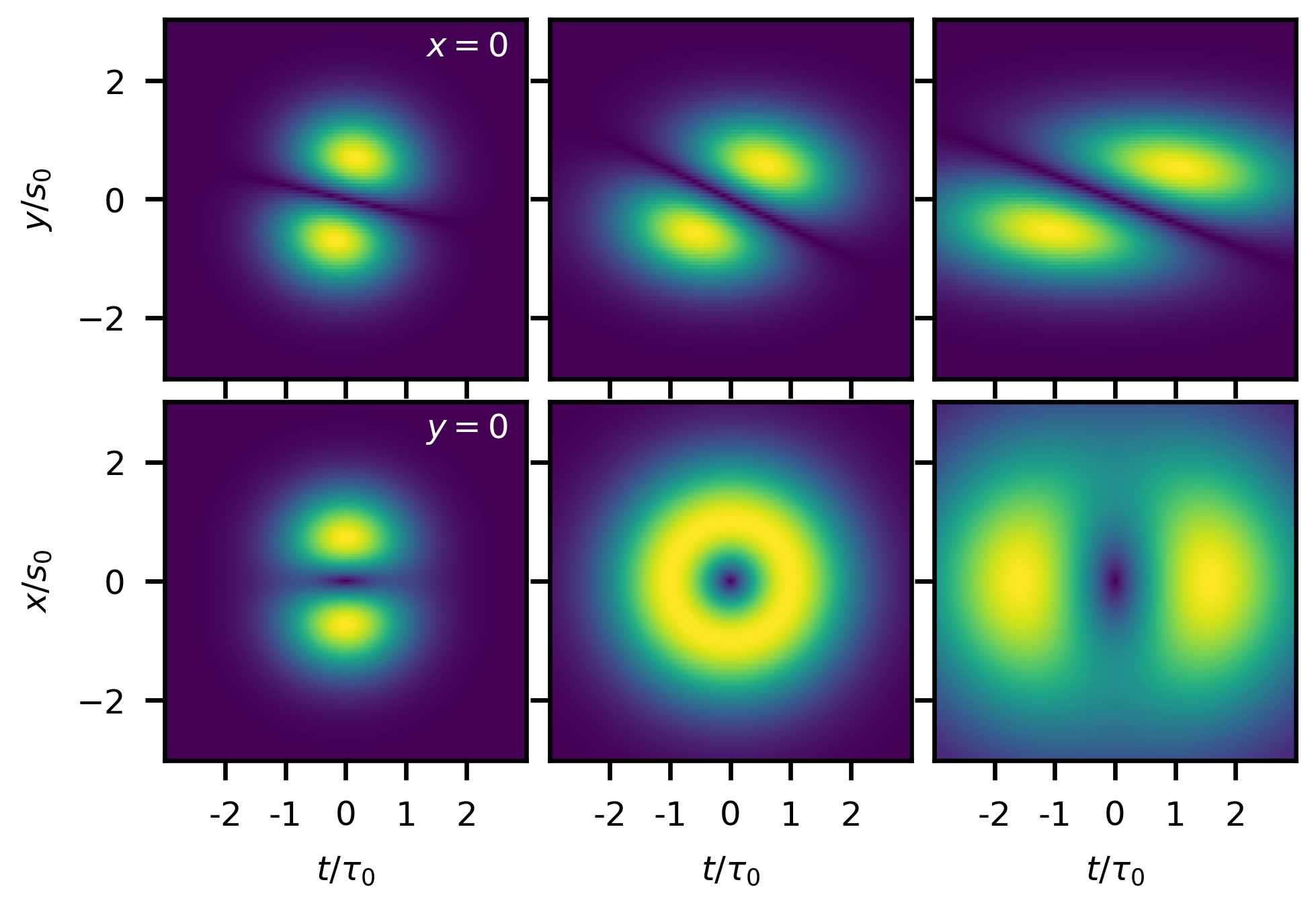}
	\caption{Amplitude of STOVs with other parameters, still at $z=0$ with $l=1$ and $b=\{0.25, 1, 2\}\times s_0/\Omega_0$ (left to right). The slope of the singularity along $y$ changes (top), and so does the amplitude distribution in the $x$-$t$ plane (bottom).}
	\label{fig:zeq0_b}
\end{figure}

When changing the value of the spatial chirp $b$, the slope $\mp (b/2\alpha^2)$ of the phase line singularity changes accordingly, taking a maximum absolute value for $b=\pm s_0/\Omega_0$, as seen in the $x=0$ sections  in Fig.~\ref{fig:zeq0_b}(top).  Increase of the absolute value of $b$ is accompanied by longer effective durations $\tau_{0,\rm eff}$ . Remarkably, a perfect donut shape in the $y=0$ sections only exists for the values $b=\pm s_0/\Omega_0$ of maximum tilt, as seen in Fig.~\ref{fig:zeq0_b}(bottom). This is because with $b=\pm s_0/\Omega_0$ the scaling in the Gaussian temporal and spatial factors are the same as the scaling in the temporal and spatial terms in the argument of the Hermite polynomial, i.e., the STOV at $y=0$ is of the form $e^{-t^{\prime 2}/\tau^{2}_{0,\rm eff}}e^{-x^2/x^{2}_{0,\rm eff}}H_l(t'/\tau_{0, \rm eff} \pm i x/s_{0,\rm eff})$, with $\tau_{0,\rm eff}=\sqrt{2}\tau_0$ and $s_{0,\rm eff}=\sqrt{2}s_0$.

\begin{figure}[htb]
	\centering
	\includegraphics[width=86mm]{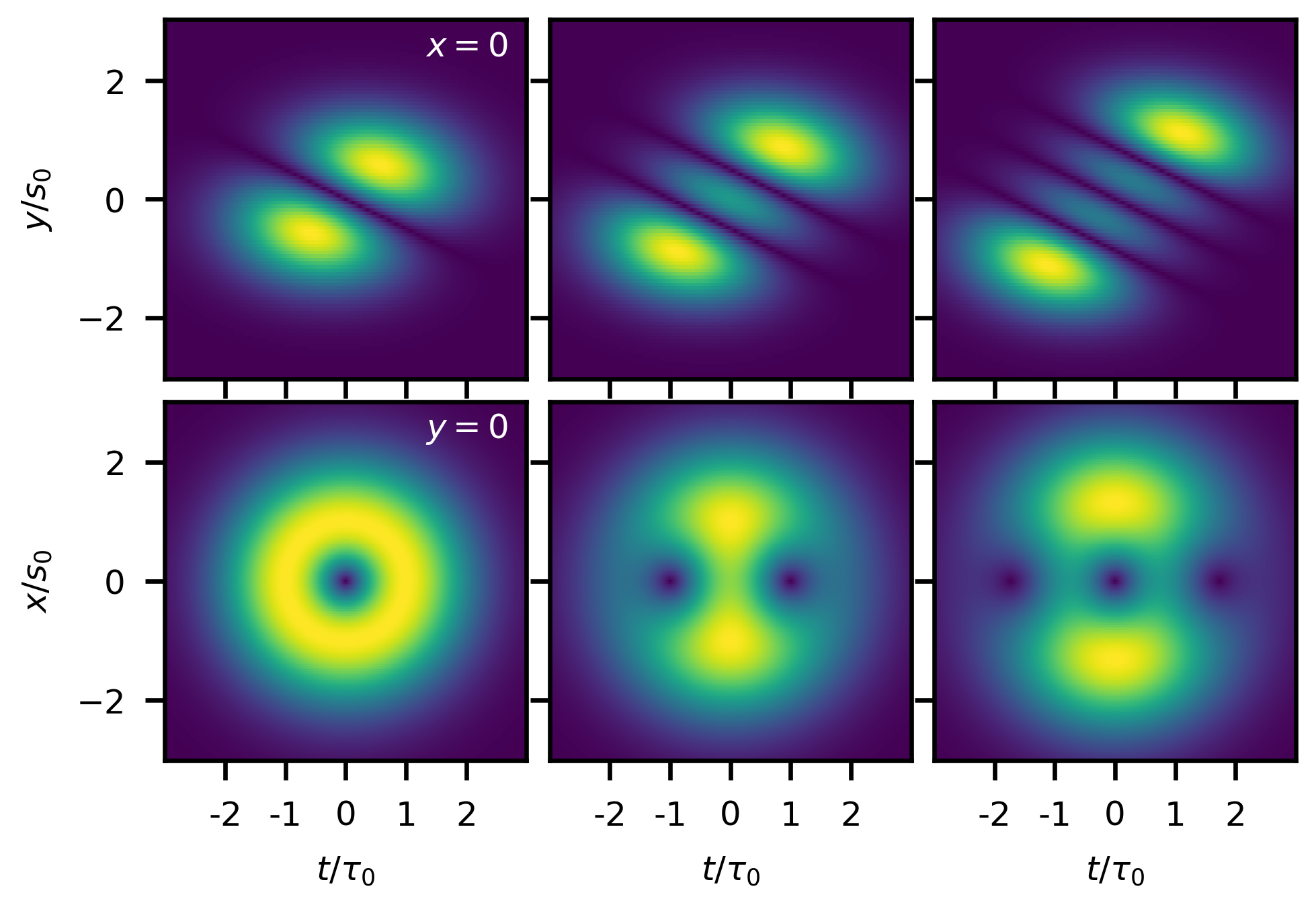}
	\caption{Amplitude of STOVs at $z=0$ with $b=s_0/\Omega_0$ and $l=1, 2$, and $3$ (left to right). There are an increasing number of line singularities in the $x=0$ plane (top), resulting in null points of amplitude in the $y=0$ plane (bottom).}
	\label{fig:zeq0_l}
\end{figure}

For general $l$, the phase singularities are located at the zeros of the Hermite polynomial, say $h_{l,n}$, with $n=1,2,\dots l$, which are all real. Thus, all phase singularities are in the plane $x=0$ and specified by the straight lines $y=\mp (b/2\alpha^2) t' \pm(b/\alpha) h_{l,n}$, being then parallel to each other with the same slope $\mp b/2\alpha^2$, as seen in Fig.~\ref{fig:zeq0_l}(top) for $l=1,2$, and $3$. In the orthogonal $y=0$ section the $l$ phase singularities manifest as $l$ null intensity points at times $t'= 2\alpha h_{l,n}$, as in Fig.~\ref{fig:zeq0_l}(bottom), which confers the intensity pattern a more complex structure as $l$ increases.

\begin{figure}[htb]
	\centering
	\includegraphics[width=86mm]{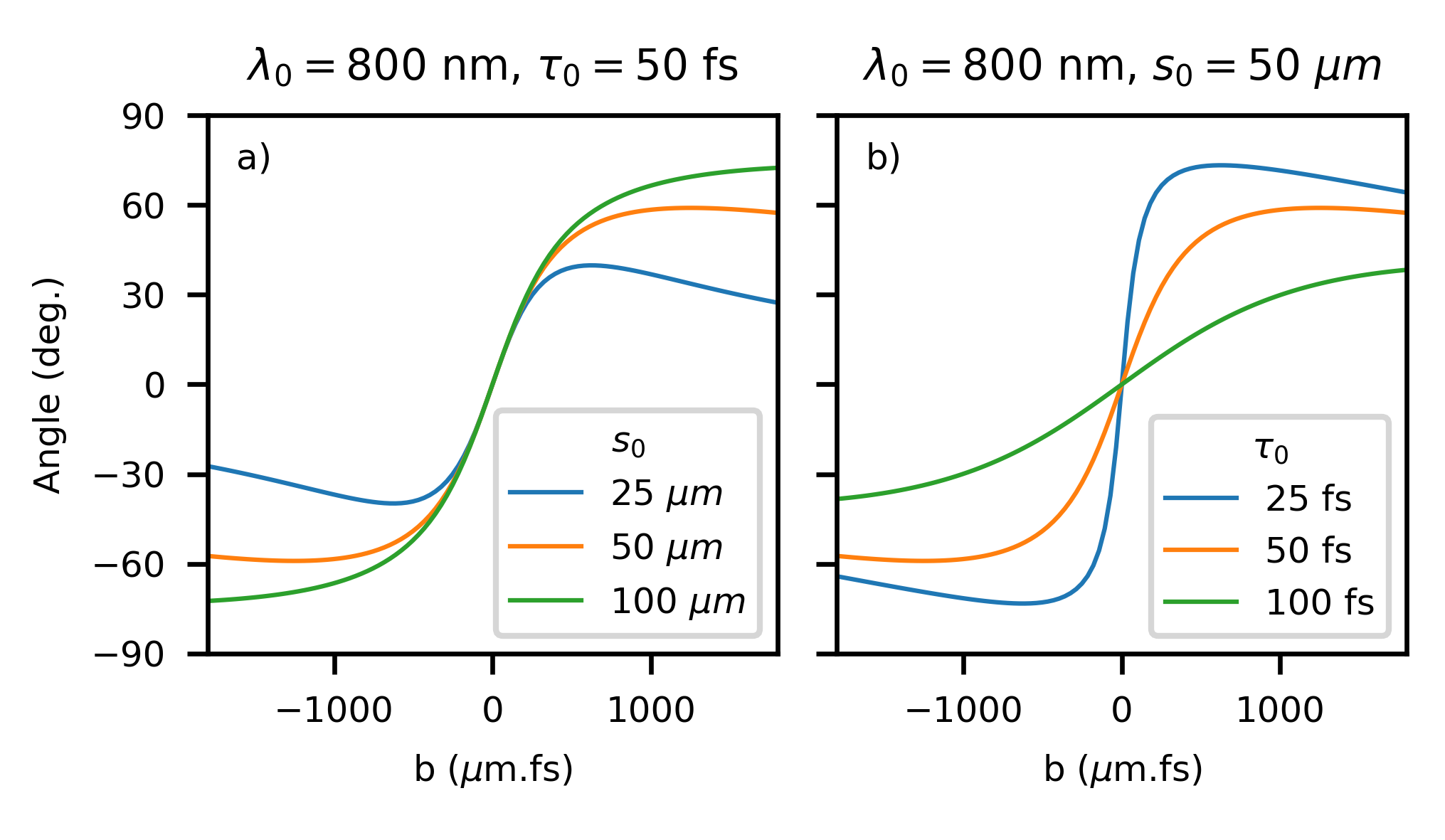}
        \includegraphics[width=43mm]{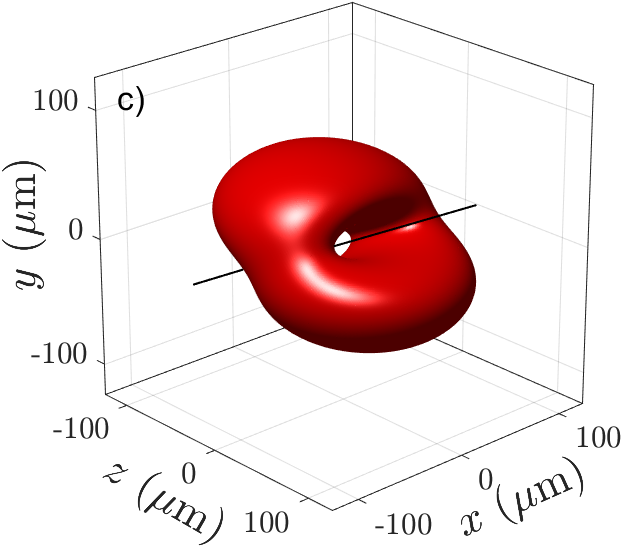}\includegraphics[width=43mm]{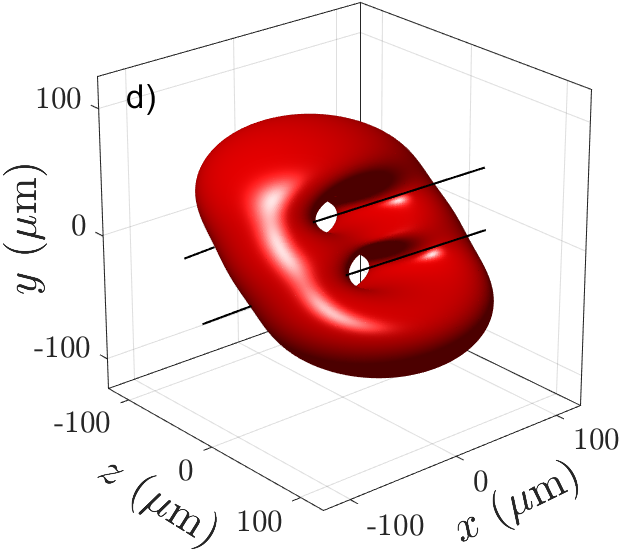}
	\caption{(a) Angle of the phase line singularity in the $z$-$y$ plane at $z=0$ as a function of $b$ for several  waist widths. The slope at $b=0$ is independent of $s_0$. (b) The same but for several pulse durations. (c,d) Snapshots ($t=0$) of iso-intensity surfaces (18\% the peak intensity) of STOVs at 800 nm, waist width $s_0=50$ $\mu$m, duration $\tau_0=100$ fs ($\Omega_0=0.02$ rad/fs), spatial chirp $b=s_0/\Omega_0$ for maximum tilt $\theta\simeq 40$ deg, and topological charges $l=1$ and $2$. The scales in $x,y$ and $z$ are the same to visualize the actual tilt.}
	\label{fig:angle}
\end{figure}

With regard to the actual orientation of the vortex line in three-dimensional space, the slope in the  $\mp b/2\alpha^2$ in the $t'$-$y$ plane amounts to a slope $\pm b/2c\alpha^2$ in the $z$-$y$ plane, or a tilt angle of the phase line singularity 
\begin{equation}\label{tilt}
\theta=\pm \tan^{-1}\left[\frac{b}{2c[(1/\Omega_0)^2+(b/s_0)^2]}\right]
\end{equation}
with respect to the $z$-axis. Figures~\ref{fig:angle} (a) and (b) illustrate the behavior of the vortex orientation depending on the spatial chirp for typical real pulse and beam parameters under paraxial and quasimonochromatic conditions, and Figs.~\ref{fig:angle} (c) and (d) show two examples of the actual tilts in space along with the three-dimensional structure of the intensity. The tilt angle is maximum for spatial chirp $b=\pm s_0/\Omega_0$, as in Fig.~\ref{fig:concept}, resulting in $\theta=\pm \tan^{-1}[s_0\Omega_0/4c]$. As seen in Fig.~\ref{fig:angle}(a), the maximum tilt angle can be as close as $90$ degrees as desired with paraxial waist widths using pulses of duration of the order of tens of femtoseconds, although $90$ degrees, i.e., a purely transversal STOV, is never reached. Evaluation of the derivative of (\ref{tilt}) with respect to $b$ at $b=0$ yields $\Omega_0^2/2c$, which is independent of $s_0$, as can be appreciated in Fig.~\ref{fig:angle}(a). Control of this derivative, and thus of the sensitivity of the tilt angle to the spatial chirp, can be exercised by the bandwidth or pulse duration, as illustrated in Fig.~\ref{fig:angle}(b). The longer the duration, the easier it is to steer the vortex in a precise direction.

As the STOV propagates from $z=0$, it diffracts as a LG beam but distorting and rotating, and maintaining the same duration due to the absence of dispersion. We could add new contour plots, but we prefer to focus on the new phenomena.  The complexity of the behavior of the singularities in purely transverse STOVs in free space and in dispersive media has been discussed only very recently~\cite{porras23-1,porras23-11}. One important conclusion is the absence of relation between topological charges and OAM except for canonical vortices (with elliptical symmetry in the case of purely transverse STOVs). The OAM is conserved on propagation, but the signs of the charges of the transverse vortices may not \cite{porras23-2}. A similar situation is observed here, affecting now to the direction of the phase line singularity, the orientation of which is seen to precess upon propagation.

\begin{figure}[htb]
\centering
\includegraphics[width=42mm]{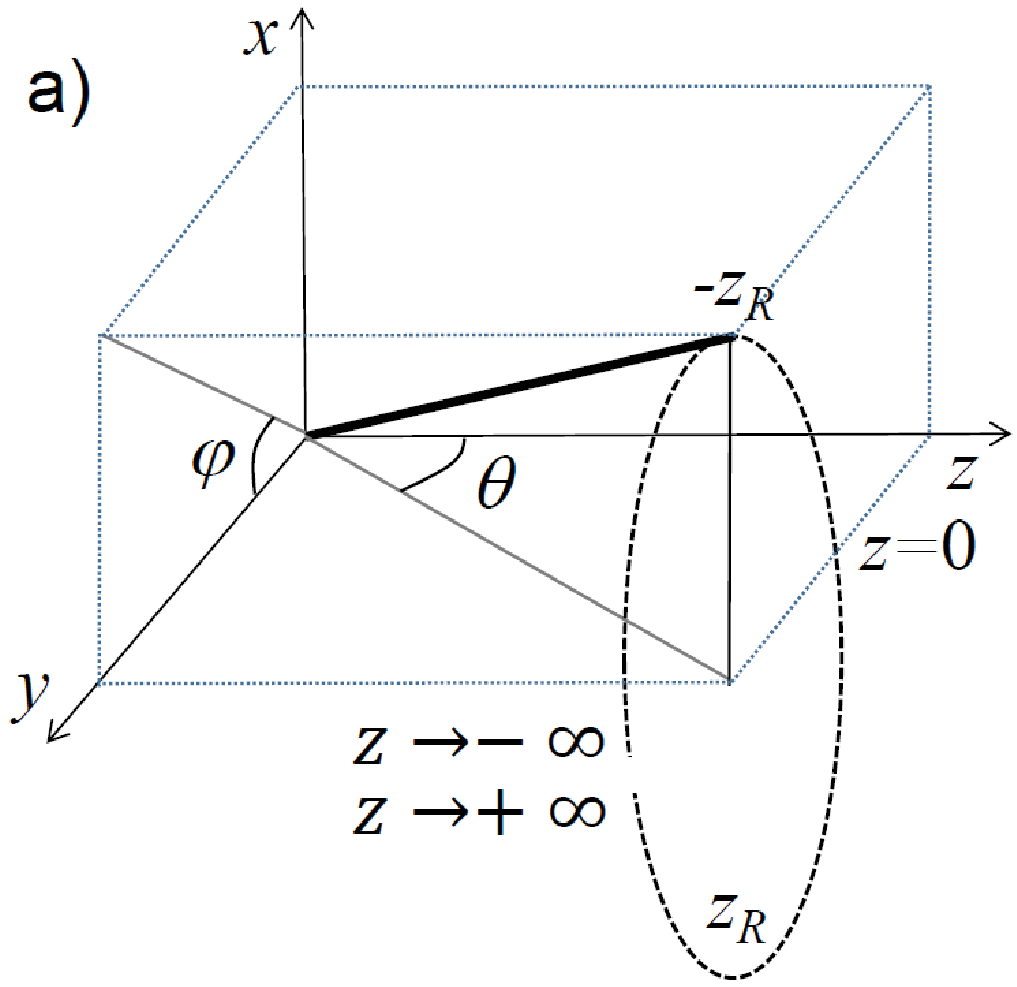}
\includegraphics[width=42mm]{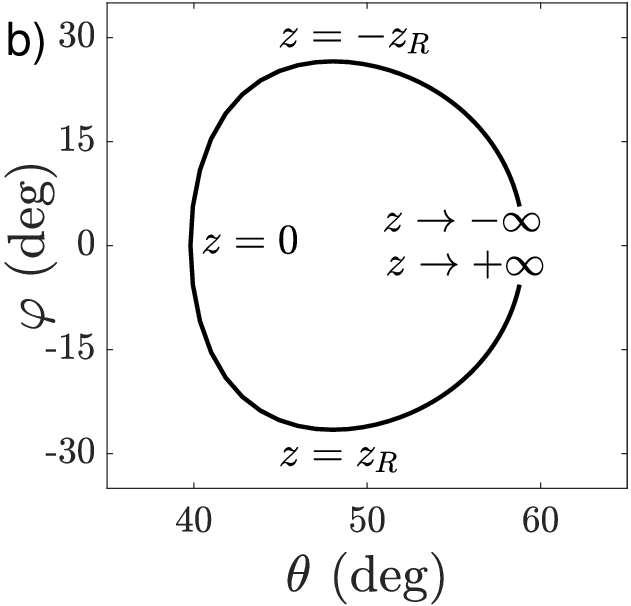}
	\caption{(a) Transveral tilt angle $\theta$ and deviation angle $\varphi$ from $x=0$. (b) Their change with $z$ for the STOV in Fig. \ref{fig:angle}(c).}
	\label{Fig6}
\end{figure}

Limiting our consideration to $l=1$, and setting the argument of the Hermite polynomial to zero in (\ref{main}), one obtains the phase line singularity at arbitrary propagation distance $z$ as the intersection of the two planes
\begin{equation}
y=\mp \frac{1}{2}\frac{b}{(1/\Omega_0)^2 +(b/s)^2}\, t',\quad x=\mp \frac{\omega_0 b^2\Omega_0^2}{2cR}\, y \,
\end{equation}
where $s$ is the Gaussian width at any distance defined above, and $1/R=z/(z^2+z_R^2)$ is the radius of curvature. From these planes, the orientation in space can be derived as above. The angle $\theta$ from the propagation direction and the new angle  of deviation from the $x=0$ plane are defined in Fig. \ref{Fig6}(a), and are plotted one as a function of the other in Fig. \ref{Fig6}(b) as the STOV in Fig. \ref{fig:angle}(c) propagates from $z=-\infty$ to $z=+\infty$.
The phase line singularity is only in the $x=0$ plane ($\varphi=0$) at the waist and at the far fields $z=\pm\infty$, with respective minimum and maximum values. The deviation angle from the $x=0$ plane is maximum at $\pm z_R$.

In conclusion, we have provided analytical expressions of fields of STOVs the phase line singularity of which can be tuned from zero to almost 90 degrees from the propagation direction. By construction, these fields can be realized experimentally by introducing a spatial chirp into a standard longitudinal vortex. The arbitrary orientation suggests that one can also tune the direction of the OAM. However, we leave for further research the amount and direction of the OAM until the current debate on the OAM oftransverse STOVs will be closed~
\cite{porras23-2,bliokh23,milchberg21}.

\section*{Funding}
Horizon 2020 Framework Programme (801505); Ministerio de Ciencia, Innovación y Universidades (PID2021-122711NB-C21).

\section*{Acknowledgments}
S.W.J. has received funding from the European Union’s Horizon 2020 research and innovation program under the Marie Skłodowska-Curie grant agreement No 801505. This work has been partially supported by the Spanish Ministry of Science and Innovation, Gobierno de España, under Contract No. PID2021-122711NB-C21.

\section*{Disclosures}
The authors declare no conflicts of interest.

\section*{Data availability}
Data underlying the results presented in this paper are not publicly available at this time but may be obtained from the authors upon reasonable request.


\begin{thebibliography}{15}%
	\makeatletter
	\providecommand \@ifxundefined [1]{%
		\@ifx{#1\undefined}
	}%
	\providecommand \@ifnum [1]{%
		\ifnum #1\expandafter \@firstoftwo
		\else \expandafter \@secondoftwo
		\fi
	}%
	\providecommand \@ifx [1]{%
		\ifx #1\expandafter \@firstoftwo
		\else \expandafter \@secondoftwo
		\fi
	}%
	\providecommand \natexlab [1]{#1}%
	\providecommand \enquote  [1]{``#1''}%
	\providecommand \bibnamefont  [1]{#1}%
	\providecommand \bibfnamefont [1]{#1}%
	\providecommand \citenamefont [1]{#1}%
	\providecommand \href@noop [0]{\@secondoftwo}%
	\providecommand \href [0]{\begingroup \@sanitize@url \@href}%
	\providecommand \@href[1]{\@@startlink{#1}\@@href}%
	\providecommand \@@href[1]{\endgroup#1\@@endlink}%
	\providecommand \@sanitize@url [0]{\catcode `\\12\catcode `\$12\catcode
		`\&12\catcode `\#12\catcode `\^12\catcode `\_12\catcode `\%12\relax}%
	\providecommand \@@startlink[1]{}%
	\providecommand \@@endlink[0]{}%
	\providecommand \url  [0]{\begingroup\@sanitize@url \@url }%
	\providecommand \@url [1]{\endgroup\@href {#1}{\urlprefix }}%
	\providecommand \urlprefix  [0]{URL }%
	\providecommand \Eprint [0]{\href }%
	\providecommand \doibase [0]{https://doi.org/}%
	\providecommand \selectlanguage [0]{\@gobble}%
	\providecommand \bibinfo  [0]{\@secondoftwo}%
	\providecommand \bibfield  [0]{\@secondoftwo}%
	\providecommand \translation [1]{[#1]}%
	\providecommand \BibitemOpen [0]{}%
	\providecommand \bibitemStop [0]{}%
	\providecommand \bibitemNoStop [0]{.\EOS\space}%
	\providecommand \EOS [0]{\spacefactor3000\relax}%
	\providecommand \BibitemShut  [1]{\csname bibitem#1\endcsname}%
	\let\auto@bib@innerbib\@empty
	%</preamble>
	\bibitem [{\citenamefont {Akturk}\ \emph {et~al.}(2010)\citenamefont {Akturk},
		\citenamefont {Gu}, \citenamefont {Bowlan},\ and\ \citenamefont
		{Trebino}}]{akturk10}%
	\BibitemOpen
	\bibfield  {author} {\bibinfo {author} {\bibfnamefont {S.}~\bibnamefont
			{Akturk}}, \bibinfo {author} {\bibfnamefont {X.}~\bibnamefont {Gu}}, \bibinfo
		{author} {\bibfnamefont {P.}~\bibnamefont {Bowlan}},\ and\ \bibinfo {author}
		{\bibfnamefont {R.}~\bibnamefont {Trebino}},\ }\bibfield  {title} {\bibinfo
		{title} {Spatio-temporal couplings in ultrashort laser pulses},\ }\href@noop
	{} {\bibfield  {journal} {\bibinfo  {journal} {Journal of Optics}\ }\textbf
		{\bibinfo {volume} {12}},\ \bibinfo {pages} {093001} (\bibinfo {year}
		{2010})}\BibitemShut {NoStop}%
	\bibitem [{\citenamefont {Shen}\ \emph {et~al.}(2023)\citenamefont {Shen},
		\citenamefont {Zhan}, \citenamefont {Wright}, \citenamefont
		{Christodoulides}, \citenamefont {Wise}, \citenamefont {Willner},
		\citenamefont {heng Zou}, \citenamefont {Zhao}, \citenamefont {Porras},
		\citenamefont {Chong}, \citenamefont {Wan}, \citenamefont {Bliokh},
		\citenamefont {Liao}, \citenamefont {Hernández-García}, \citenamefont
		{Murnane}, \citenamefont {Yessenov}, \citenamefont {Abouraddy}, \citenamefont
		{Wong}, \citenamefont {Go}, \citenamefont {Kumar}, \citenamefont {Guo},
		\citenamefont {Fan}, \citenamefont {Papasimakis}, \citenamefont {Zheludev},
		\citenamefont {Chen}, \citenamefont {Zhu}, \citenamefont {Agrawal},
		\citenamefont {Mounaix}, \citenamefont {Fontaine}, \citenamefont {Carpenter},
		\citenamefont {Jolly}, \citenamefont {Dorrer}, \citenamefont {Alonso},
		\citenamefont {Lopez-Quintas}, \citenamefont {López-Ripa}, \citenamefont
		{Íñigo J.~Sola}, \citenamefont {Huang}, \citenamefont {Zhang},
		\citenamefont {Ruan}, \citenamefont {Dorrah}, \citenamefont {Capasso},\ and\
		\citenamefont {Forbes}}]{shenY23}%
	\BibitemOpen
	\bibfield  {author} {\bibinfo {author} {\bibfnamefont {Y.}~\bibnamefont
			{Shen}}, \bibinfo {author} {\bibfnamefont {Q.}~\bibnamefont {Zhan}}, \bibinfo
		{author} {\bibfnamefont {L.~G.}\ \bibnamefont {Wright}}, \bibinfo {author}
		{\bibfnamefont {D.~N.}\ \bibnamefont {Christodoulides}}, \bibinfo {author}
		{\bibfnamefont {F.~W.}\ \bibnamefont {Wise}}, \bibinfo {author}
		{\bibfnamefont {A.~E.}\ \bibnamefont {Willner}}, \bibinfo {author}
		{\bibfnamefont {K.}~\bibnamefont {heng Zou}}, \bibinfo {author}
		{\bibfnamefont {Z.}~\bibnamefont {Zhao}}, \bibinfo {author} {\bibfnamefont
			{M.~A.}\ \bibnamefont {Porras}}, \bibinfo {author} {\bibfnamefont
			{A.}~\bibnamefont {Chong}}, \bibinfo {author} {\bibfnamefont
			{C.}~\bibnamefont {Wan}}, \bibinfo {author} {\bibfnamefont {K.~Y.}\
			\bibnamefont {Bliokh}}, \bibinfo {author} {\bibfnamefont {C.-T.}\
			\bibnamefont {Liao}}, \bibinfo {author} {\bibfnamefont {C.}~\bibnamefont
			{Hernández-García}}, \bibinfo {author} {\bibfnamefont {M.~M.}\ \bibnamefont
			{Murnane}}, \bibinfo {author} {\bibfnamefont {M.}~\bibnamefont {Yessenov}},
		\bibinfo {author} {\bibfnamefont {A.~F.}\ \bibnamefont {Abouraddy}}, \bibinfo
		{author} {\bibfnamefont {L.~J.}\ \bibnamefont {Wong}}, \bibinfo {author}
		{\bibfnamefont {M.}~\bibnamefont {Go}}, \bibinfo {author} {\bibfnamefont
			{S.}~\bibnamefont {Kumar}}, \bibinfo {author} {\bibfnamefont
			{C.}~\bibnamefont {Guo}}, \bibinfo {author} {\bibfnamefont {S.}~\bibnamefont
			{Fan}}, \bibinfo {author} {\bibfnamefont {N.}~\bibnamefont {Papasimakis}},
		\bibinfo {author} {\bibfnamefont {N.~I.}\ \bibnamefont {Zheludev}}, \bibinfo
		{author} {\bibfnamefont {L.}~\bibnamefont {Chen}}, \bibinfo {author}
		{\bibfnamefont {W.}~\bibnamefont {Zhu}}, \bibinfo {author} {\bibfnamefont
			{A.}~\bibnamefont {Agrawal}}, \bibinfo {author} {\bibfnamefont
			{M.}~\bibnamefont {Mounaix}}, \bibinfo {author} {\bibfnamefont {N.~K.}\
			\bibnamefont {Fontaine}}, \bibinfo {author} {\bibfnamefont {J.}~\bibnamefont
			{Carpenter}}, \bibinfo {author} {\bibfnamefont {S.~W.}\ \bibnamefont
			{Jolly}}, \bibinfo {author} {\bibfnamefont {C.}~\bibnamefont {Dorrer}},
		\bibinfo {author} {\bibfnamefont {B.}~\bibnamefont {Alonso}}, \bibinfo
		{author} {\bibfnamefont {I.}~\bibnamefont {Lopez-Quintas}}, \bibinfo {author}
		{\bibfnamefont {M.}~\bibnamefont {López-Ripa}}, \bibinfo {author}
		{\bibnamefont {Íñigo J.~Sola}}, \bibinfo {author} {\bibfnamefont
			{J.}~\bibnamefont {Huang}}, \bibinfo {author} {\bibfnamefont
			{H.}~\bibnamefont {Zhang}}, \bibinfo {author} {\bibfnamefont
			{Z.}~\bibnamefont {Ruan}}, \bibinfo {author} {\bibfnamefont {A.~H.}\
			\bibnamefont {Dorrah}}, \bibinfo {author} {\bibfnamefont {F.}~\bibnamefont
			{Capasso}},\ and\ \bibinfo {author} {\bibfnamefont {A.}~\bibnamefont
			{Forbes}},\ }\bibfield  {title} {\bibinfo {title} {Roadmap on spatiotemporal
			light fields},\ }\href@noop {} {\bibfield  {journal} {\bibinfo  {journal}
			{Journal of Optics}\ }\textbf {\bibinfo {volume} {accepted}} (\bibinfo {year}
		{2023})}\BibitemShut {NoStop}%
	\bibitem [{\citenamefont {Jhajj}\ \emph {et~al.}(2016)\citenamefont {Jhajj},
		\citenamefont {Larkin}, \citenamefont {Rosenthal}, \citenamefont {Zahedpour},
		\citenamefont {Wahlstrand},\ and\ \citenamefont {Milchberg}}]{jhajj16}%
	\BibitemOpen
	\bibfield  {author} {\bibinfo {author} {\bibfnamefont {N.}~\bibnamefont
			{Jhajj}}, \bibinfo {author} {\bibfnamefont {I.}~\bibnamefont {Larkin}},
		\bibinfo {author} {\bibfnamefont {E.~W.}\ \bibnamefont {Rosenthal}}, \bibinfo
		{author} {\bibfnamefont {S.}~\bibnamefont {Zahedpour}}, \bibinfo {author}
		{\bibfnamefont {J.~K.}\ \bibnamefont {Wahlstrand}},\ and\ \bibinfo {author}
		{\bibfnamefont {H.~M.}\ \bibnamefont {Milchberg}},\ }\bibfield  {title}
	{\bibinfo {title} {Spatiotemporal optical vortices},\ }\href@noop {}
	{\bibfield  {journal} {\bibinfo  {journal} {Physical Review X}\ }\textbf
		{\bibinfo {volume} {6}},\ \bibinfo {pages} {031037} (\bibinfo {year}
		{2016})}\BibitemShut {NoStop}%
	\bibitem [{\citenamefont {Hancock}\ \emph {et~al.}(2019)\citenamefont
		{Hancock}, \citenamefont {Zahedpour}, \citenamefont {Goffin},\ and\
		\citenamefont {Milchberg}}]{hancock19}%
	\BibitemOpen
	\bibfield  {author} {\bibinfo {author} {\bibfnamefont {S.~W.}\ \bibnamefont
			{Hancock}}, \bibinfo {author} {\bibfnamefont {S.}~\bibnamefont {Zahedpour}},
		\bibinfo {author} {\bibfnamefont {A.}~\bibnamefont {Goffin}},\ and\ \bibinfo
		{author} {\bibfnamefont {H.~M.}\ \bibnamefont {Milchberg}},\ }\bibfield
	{title} {\bibinfo {title} {Free-space propagation of spatiotemporal optical
			vortices},\ }\href@noop {} {\bibfield  {journal} {\bibinfo  {journal}
			{Optica}\ }\textbf {\bibinfo {volume} {6}},\ \bibinfo {pages} {1547}
		(\bibinfo {year} {2019})}\BibitemShut {NoStop}%
	\bibitem [{\citenamefont {Chong}\ \emph {et~al.}(2020)\citenamefont {Chong},
		\citenamefont {Wan}, \citenamefont {Chen},\ and\ \citenamefont
		{Zhan}}]{chong20}%
	\BibitemOpen
	\bibfield  {author} {\bibinfo {author} {\bibfnamefont {A.}~\bibnamefont
			{Chong}}, \bibinfo {author} {\bibfnamefont {C.}~\bibnamefont {Wan}}, \bibinfo
		{author} {\bibfnamefont {J.}~\bibnamefont {Chen}},\ and\ \bibinfo {author}
		{\bibfnamefont {Q.}~\bibnamefont {Zhan}},\ }\bibfield  {title} {\bibinfo
		{title} {Generation of spatiotemporal optical vortices with controllable
			transverse orbital angular momentum},\ }\href@noop {} {\bibfield  {journal}
		{\bibinfo  {journal} {Nature Photonics}\ }\textbf {\bibinfo {volume} {14}},\
		\bibinfo {pages} {350} (\bibinfo {year} {2020})}\BibitemShut {NoStop}%
	\bibitem [{\citenamefont {Wang}\ \emph {et~al.}(2021)\citenamefont {Wang},
		\citenamefont {Guo}, \citenamefont {Jin}, \citenamefont {Song},\ and\
		\citenamefont {Fan}}]{wangH21}%
	\BibitemOpen
	\bibfield  {author} {\bibinfo {author} {\bibfnamefont {H.}~\bibnamefont
			{Wang}}, \bibinfo {author} {\bibfnamefont {C.}~\bibnamefont {Guo}}, \bibinfo
		{author} {\bibfnamefont {W.}~\bibnamefont {Jin}}, \bibinfo {author}
		{\bibfnamefont {A.~Y.}\ \bibnamefont {Song}},\ and\ \bibinfo {author}
		{\bibfnamefont {S.}~\bibnamefont {Fan}},\ }\bibfield  {title} {\bibinfo
		{title} {Engineering arbitrarily oriented spatiotemporal optical vortices
			using transmission nodal lines},\ }\href@noop {} {\bibfield  {journal}
		{\bibinfo  {journal} {Optica}\ }\textbf {\bibinfo {volume} {8}},\ \bibinfo
		{pages} {966} (\bibinfo {year} {2021})}\BibitemShut {NoStop}%
	\bibitem [{\citenamefont {Zang}\ \emph {et~al.}(2022)\citenamefont {Zang},
		\citenamefont {Mirando},\ and\ \citenamefont {Chong}}]{zangY22}%
	\BibitemOpen
	\bibfield  {author} {\bibinfo {author} {\bibfnamefont {Y.}~\bibnamefont
			{Zang}}, \bibinfo {author} {\bibfnamefont {A.}~\bibnamefont {Mirando}},\ and\
		\bibinfo {author} {\bibfnamefont {A.}~\bibnamefont {Chong}},\ }\bibfield
	{title} {\bibinfo {title} {Spatiotemporal optical vortices with arbitrary
			orbital angular momentum orientation by astigmatic mode converters},\
	}\href@noop {} {\bibfield  {journal} {\bibinfo  {journal} {Nanophotonics}\
		}\textbf {\bibinfo {volume} {11}},\ \bibinfo {pages} {745} (\bibinfo {year}
		{2022})}\BibitemShut {NoStop}%
	\bibitem [{\citenamefont {Talposi}\ \emph {et~al.}(2022)\citenamefont
		{Talposi}, \citenamefont {Iancu},\ and\ \citenamefont {Ursescu}}]{talposi22}%
	\BibitemOpen
	\bibfield  {author} {\bibinfo {author} {\bibfnamefont {A.-M.}\ \bibnamefont
			{Talposi}}, \bibinfo {author} {\bibfnamefont {V.}~\bibnamefont {Iancu}},\
		and\ \bibinfo {author} {\bibfnamefont {D.}~\bibnamefont {Ursescu}},\
	}\bibfield  {title} {\bibinfo {title} {Influence of spatio-temporal couplings
			on focused optical vortices},\ }\href@noop {} {\bibfield  {journal} {\bibinfo
			{journal} {Photonics}\ }\textbf {\bibinfo {volume} {9}},\ \bibinfo {pages}
		{389} (\bibinfo {year} {2022})}\BibitemShut {NoStop}%
	\bibitem [{\citenamefont {Gu}\ \emph {et~al.}(2004)\citenamefont {Gu},
		\citenamefont {Akturk},\ and\ \citenamefont {Trebino}}]{gu04}%
	\BibitemOpen
	\bibfield  {author} {\bibinfo {author} {\bibfnamefont {X.}~\bibnamefont
			{Gu}}, \bibinfo {author} {\bibfnamefont {S.}~\bibnamefont {Akturk}},\ and\
		\bibinfo {author} {\bibfnamefont {R.}~\bibnamefont {Trebino}},\ }\bibfield
	{title} {\bibinfo {title} {Spatial chirp in ultrafast optics},\ }\href@noop
	{} {\bibfield  {journal} {\bibinfo  {journal} {Optics Communications}\
		}\textbf {\bibinfo {volume} {242}},\ \bibinfo {pages} {599} (\bibinfo {year}
		{2004})}\BibitemShut {NoStop}%
	\bibitem [{\citenamefont {Gradshteyn}\ and\ \citenamefont
		{Ryzhik}(2007)}]{gradshteyn07}%
	\BibitemOpen
	\bibfield  {author} {\bibinfo {author} {\bibfnamefont {I.~S.}\ \bibnamefont
			{Gradshteyn}}\ and\ \bibinfo {author} {\bibfnamefont {I.~M.}\ \bibnamefont
			{Ryzhik}},\ }\href@noop {} {\emph {\bibinfo {title} {Table of Integrals,
				series, and products}}},\ \bibinfo {edition} {seventh}\ ed.\ (\bibinfo
	{publisher} {Academic Press},\ \bibinfo {year} {2007})\BibitemShut {NoStop}%
	\bibitem [{\citenamefont {Porras}(2023{\natexlab{a}})}]{porras23-1}%
	\BibitemOpen
	\bibfield  {author} {\bibinfo {author} {\bibfnamefont {M.~A.}\ \bibnamefont
			{Porras}},\ }\bibfield  {title} {\bibinfo {title} {Propagation of
			higher-order spatiotemporal vortices},\ }\href@noop {} {\bibfield  {journal}
		{\bibinfo  {journal} {Optics Letters}\ }\textbf {\bibinfo {volume} {48}},\
		\bibinfo {pages} {367} (\bibinfo {year} {2023}{\natexlab{a}})}\BibitemShut
	{NoStop}%
	\bibitem [{\citenamefont {Porras}(2023{\natexlab{b}})}]{porras23-11}%
	\BibitemOpen
	\bibfield  {author} {\bibinfo {author} {\bibfnamefont {M.~A.}\ \bibnamefont
			{Porras}},\ }\bibfield  {title} {\bibinfo {title} {Propagation of
			spatiotemporal optical vortex beams in linear, second-order dispersive
			media},\ }\href@noop {} {\bibfield  {journal} {\bibinfo  {journal} {Physical
				Review A}\ }\textbf {\bibinfo {volume} {accepted}} (\bibinfo {year}
		{2023}{\natexlab{b}})}\BibitemShut {NoStop}%
	\bibitem [{\citenamefont {Porras}(2023{\natexlab{c}})}]{porras23-2}%
	\BibitemOpen
	\bibfield  {author} {\bibinfo {author} {\bibfnamefont {M.~A.}\ \bibnamefont
			{Porras}},\ }\bibfield  {title} {\bibinfo {title} {Transverse orbital angular
			momentum of spatiotemporal optical vortices},\ }\href@noop {} {\bibfield
		{journal} {\bibinfo  {journal} {Progress in Electromagnetic Research}\
		}\textbf {\bibinfo {volume} {177}},\ \bibinfo {pages} {95} (\bibinfo {year}
		{2023}{\natexlab{c}})}\BibitemShut {NoStop}%
	\bibitem [{\citenamefont {Bliokh}(2023)}]{bliokh23}%
	\BibitemOpen
	\bibfield  {author} {\bibinfo {author} {\bibfnamefont {K.~Y.}\ \bibnamefont
			{Bliokh}},\ }\bibfield  {title} {\bibinfo {title} {Orbital angular momentum
			of optical, acoustic, and quantum-mechanical spatiotemporal vortex pulses},\
	}\href@noop {} {\bibfield  {journal} {\bibinfo  {journal} {Physical Review
				A}\ }\textbf {\bibinfo {volume} {107}},\ \bibinfo {pages} {L031501} (\bibinfo
		{year} {2023})}\BibitemShut {NoStop}%
	\bibitem [{\citenamefont {Hancock}\ \emph {et~al.}(2021)\citenamefont
		{Hancock}, \citenamefont {Zahedpour},\ and\ \citenamefont
		{Milchberg}}]{milchberg21}%
	\BibitemOpen
	\bibfield  {author} {\bibinfo {author} {\bibfnamefont {S.~W.}\ \bibnamefont
			{Hancock}}, \bibinfo {author} {\bibfnamefont {S.}~\bibnamefont {Zahedpour}},\
		and\ \bibinfo {author} {\bibfnamefont {H.~M.}\ \bibnamefont {Milchberg}},\
	}\bibfield  {title} {\bibinfo {title} {Mode structure and orbital angular
			momentum of spatiotemporal optical vortex pulses},\ }\href@noop {} {\bibfield
		{journal} {\bibinfo  {journal} {Physical Review Letter}\ }\textbf {\bibinfo
			{volume} {127}},\ \bibinfo {pages} {193901} (\bibinfo {year}
		{2021})}\BibitemShut {NoStop}%
\end{thebibliography}
\end{document}